\documentclass[preprint,pre,showpacs,keywords]{revtex4}
\usepackage{epsfig}
\usepackage{amsmath}
\begin{document}
\title{Chaotic advection of inertial particles in two dimensional flows}
\author{Neelima Gupte}
\email{gupte@physics.iitm.ac.in}
\author{N. Nirmal Thyagu}
\email{nirmal@physics.iitm.ac.in}
\affiliation{Department of Physics, Indian Institute of Technology Madras, Chennai-600036,India}
\keywords{Bailout embedding, Bifurcation diagram, Attractor merging crisis}
\date{\today}
\pacs{05.45, 47.52.+j}
\begin{abstract}

\indent We study the dynamics of inertial particles in two dimensional
incompressible flows.  
The Maxey-Riley equation describing the motion
of inertial particles is used to construct a four dimensional 
dissipative bailout embedding map. This map models the dynamics
of the inertial particles while the base flow is represented by a
$2-d$ area preserving map. The dynamics of particles heavier
than the fluid, the aerosols, as well as that of bubbles, particles lighter than the 
fluid, can be classified into 3 main dynamical 
regimes - periodic orbits, chaotic structures and mixed regions.
A phase diagram in the parameter space is constructed with 
the Lyapunov characteristic exponents of the $4-d$ map in which 
these dynamical regimes are distinctly identified. 
The embedding map can target periodic orbits, as well as chaotic
structures, in both the aerosol and bubble regimes, 
at suitable  values of the dissipation parameter.
\end{abstract}

\maketitle

\section{Introduction}
	
The motion of inertial particles in fluids is governed 
by dynamical equations which display rich and 
complex behaviour. The inertial particles have 
density that is different from that of the fluid
in which they are immersed, in contrast to the neutral
particles which have the same density of
the fluid. The dynamics of the inertial particles 
deserve attention from both the point of view of fundamental physics 
as they  exhibit complex behaviour, as well as
from that of their 
applicability to practical situations e.g. 
the transport of pollutants in the atmosphere
and plankton in oceans \cite{mot03,reig01,babia00,cart02,benc03,raf06}. 

If impurities in fluids are modeled by small spherical tracers, 
the Lagrangian dynamics of  such small spherical tracers in  two dimensional
incompressible fluid flows is described by the Maxey-Riley equations.
These are further simplified under various approximations to give a 
set of minimal equations called the embedding equations where the fluid
flow dynamics is embedded in a larger set of equations which include 
the differences between the particle and fluid
velocities\cite{mot03,cart02}. 
Although the Lagrangian dynamics of the underlying fluid flow is
incompressible, the particle motion is  compressible \cite{maxey87},    
and has regions of contraction and expansion. The density
grows in the former giving rise to clusters and
falls in the latter giving rise to voids. The properties of the base        
flow have important consequences for the transport and mixing of
particles. Map analogs of the embedding equations have also been 
constructed for cases where the fluid dynamics is modeled by
area-preserving maps which essentially retain the qualitative
features of the flow \cite{pierrehumbert00,fereday02}.
The embedded  dynamics in both cases is dissipative in nature.

Further complexity is added to the dynamics by the density difference
between the particles and the fluid. In the case of two-dimensional
chaotic flows, it has been observed earlier that particles
with density higher than the base flow, the aerosols,
tend to migrate away from the KAM islands, while particles
lighter than the fluid, the bubbles, display the opposite tendency
\cite{cartprl02}. Neutrally buoyant particles also showed a similar
result,  wherein the particles settled into         
the KAM islands. Our study indicates that in addition to the density 
difference, the dissipation parameter of the system also has
an important role to play in the dynamic behaviour of the aerosols and       
the bubbles.           

In Sec. II of  this paper, the Maxey-Riley 
equation describing the motion of inertial particles
is simplified to get a minimal equation of motion.
A map analog of this minimal equation  called the embedding map,
is obtained to model the particle dynamics in discrete time. 
The fluid dynamics is represented by a $2-d$ base map, which we
choose to be the standard map.
The  embedding map is four dimensional and 
dissipative, while the base map is  two dimensional
and area-preserving. Unlike the earlier results mentioned 
above, viz. that the bubbles tended to form structures 
in the KAM islands\cite{cartprl02} and the aerosols
were pushed away, we found that structures form
for both bubbles and aerosols in certain parameter
regimes due to the role of the dissipation parameter. 
Both the aerosol and bubble regimes in the phase diagram
show regions where periodic orbits as well as  
chaotic  structures can be seen in the phase space plots.
In Sec. III we obtain  the phase diagram of the system
in the $\alpha-\gamma$ space where $\alpha$ is the 
mass ratio parameter, and $\gamma$ is the dissipation parameter. 
The phase diagram shows rich structure in the $\alpha < 1$
aerosol regime, as well as the $\alpha > 1$ bubble
regime. In addition to these, fully or partially mixed regimes
can also be seen in  both the aerosol and bubble cases.
Thus the dynamic behaviour of the inertial particles can be of  
three major types. The Lyapunov exponents of the four dimensional
map is used as the characterizers in demarcating these 
three major regimes. Our results can have implications for 
practical problems such as the dispersion of pollutants 
by atmospheric flows, and catalytic chemical reactions.

\section{The Embedding Equation}

The motion of small spherical particles in an incompressible 
fluid has been studied extensively. The basic equation of
motion was derived by Maxey and Riley \cite{maxey83},
and is given by \cite{babia00},

\begin{equation}
\begin{split}
\rho_{p} \frac{d{\bf v}}{dt} & =  \rho_{f} \frac{D{\bf u}}{Dt} + (\rho_{p}-\rho_{f}) {\bf g}\\
&\quad - \frac{9 \nu \rho_{f}}{2a^2}\left({\bf v}-{\bf u}-\frac{a^2}{6} \nabla^{2} {\bf u}\right)\\
&\quad - \frac{\rho_{f}}{2}\left(\frac{d {\bf v}}{dt} - \frac{D}{Dt}\left[{\bf u} + \frac{a^{2}}{10} \nabla^{2} {\bf u}\right]\right)\\
&\quad - \frac{9 \rho_{f}}{2a} \sqrt{\frac{\nu}{\pi}} \int^t_0  \frac{1}{\sqrt{t-\xi}}\frac{d}{d\xi}({\bf v}-{\bf u}-\frac{a^{2}}{6}\nabla^{2}{\bf u}) \mathrm{d}\xi .
\end{split}
\label{maxey}
\end{equation}

Here ${\bf v}$ represents the particle velocity, ${\bf u}$ the fluid
velocity, $\rho_{p}$ the density of the particle, $\rho_{f}$ the density of
the fluid, and $\nu$, $a$, ${\bf g}$ represent the kinematic viscosity  of the fluid, the radius of the
particle and  the acceleration due to gravity respectively.

The first  term in the right of Eqn. \ref{maxey} represents the force exerted by the
undisturbed flow on the particle, the second term represents the buoyancy,
the third term represents the  Stokes drag, the fourth term represents the  added mass, and
the last the Basset-Bossinesq history force term. The derivative $D{\bf u}/Dt$ is taken along
the path of the fluid element, $ D{\bf u}/Dt = {\partial u}/{\partial t} + ({\bf
u} \cdot \nabla) {\bf u}$. The derivative $d{\bf u}/dt$, is taken along the 
trajectory of the particle $d{\bf u}/dt = {\partial {\bf u}}/{\partial t} +
({\bf v} \cdot \nabla) {\bf u}$.

In deriving Eqn. \ref{maxey} it was assumed\cite{maxey83,babia00} that
the particle radius, and the  Reynolds number are
small as well as the velocity gradients around the particle. It is also
assumed  that the initial velocities
of the particle and the fluid are same. A full review of the problem can be
found in Ref. \cite{michael97}.

Now Eqn. \ref{maxey} can be simplified through the following arguments. The Faxen
correction $a^{2} \nabla^{2} {\bf u}$ is of the magnitude $O(a^{2}u/L)$, and
from the assumption, $a<<L$, this term's contribution 
becomes negligible and can be excluded from the equation.
The Basset history force term which takes into account 
viscous memory effects becomes less significant and can be dropped,
as  the particle size is sufficiently small and the concentration of
particles is
sufficiently low, 
that they do not modify the flow field or interact with each other 
\cite{benc03,michael97}. Under the low Reynolds
number approximation , both the derivatives $D{\bf u}/Dt$ and $d{\bf v}/dt$
 will approximately be the same \cite{maxey83}. We assume the buoyancy effects
to be negligible. We thus arrive at the following equation,

\begin{eqnarray}
\rho_{p} \frac{d {\bf v}}{dt} &=& \rho_{f} \frac{d {\bf u}}{dt} - \frac{9 \nu
\rho_{f}}{2a^{2}} ({\bf v - u}) -\frac{\rho_{f}}{2}\left(\frac{d {\bf v}}{dt} - \frac{d{\bf u}}{dt}\right).
\label{simp_one}
\end{eqnarray}

This can be easily written into a non-dimensional form by proper rescaling of
length, time and velocity by scale factors $L, T = L/U$ ,and $U$ respectively, to obtain,

\begin{eqnarray}
\frac{d{\bf v}}{dt} - \alpha \frac{d{\bf u}}{dt} & = &- \frac{2}{3} \left(\frac{9 \alpha }{2 a^2 Re} \right) ({\bf v-u}),
\label{simp_two}
\end{eqnarray}

where $\alpha = 3 \rho_{f}/(2 \rho_{p}+\rho_{f})$ is the mass ratio parameter,
$Re = (UL)/ \nu$, is the Reynolds number of the fluid. 
By using the particle Stokes number $St = (2/9)a^2 Re$ and 
defining the dissipation parameter as $\gamma = (2\alpha/3 St)$, the final
simplified equation takes the form,

\begin{eqnarray}
\frac{d{\bf v}}{dt} - \alpha \frac{d{\bf u}}{dt}  & = & -\gamma({\bf v - u}).
\label{particle}
\end{eqnarray}

\begin{figure}[!t]
	\begin{center}
	\resizebox{65mm}{65mm}{\includegraphics{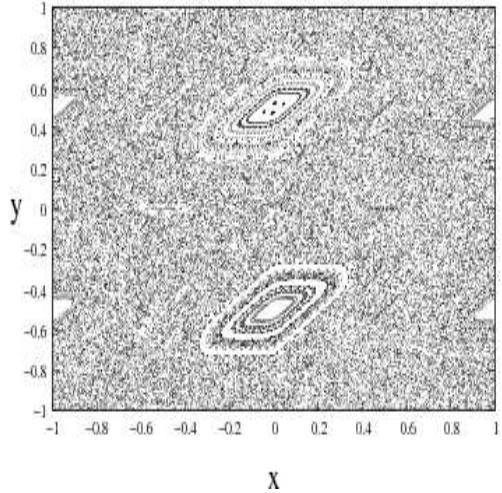}}\\
	\end{center}
  \caption{ The phase space plot of the standard map for $K=2.0$. 
Random initial conditions distributed uniformly in the phase space were
chosen.} 
\label{phsp-std}
\end{figure}
                    
The map analog of Eqn. \ref{particle} can be constructed. 
Let the base flow be represented by  an area preserving map $M$ 
with the map evolution equation ${\bf x}_{n+1} =  {\bf M}({\bf x}_{n})$.
As the particle dynamics in the fluid flow is compressible, it can 
be represented by a dissipative map. The map analog of Eqn. \ref{particle}
has been chosen in \cite{mot03} to  be,

\begin{eqnarray}
{\bf x}_{n+2} - {\bf M}({\bf x}_{n+1}) &  = & e^{-\gamma} (\alpha {\bf x}_{n+1} - {\bf M}({\bf x}_n)).
\end{eqnarray}

This equation can be rewritten by introducing a  new variable $\delta$, 
which defines the detachment of the particle from the local 
fluid parcel as,
\begin{eqnarray}
{\bf x}_{n+1} & = & {\bf M(x}_n)+{\bf \delta}_n \nonumber \\
{\bf {\delta}}_{n+1} & = & e^{-\gamma}[\alpha {\bf x}_{n+1}-{\bf M(x}_n)].
\end{eqnarray}

 This is the bailout embedding map. When the detachment measured by
 the $\delta$ is nonzero, the particle is said to have bailed out of the fluid 
trajectory. In the limit $\gamma \to \infty$ and $\alpha = 1$, ${\bf \delta} 
\to 0$ and the fluid dynamics is recovered. So the fluid dynamics is embedded 
in the particle's dynamics and is recovered under appropriate limit. This map
is dissipative with the phase space contraction rate  being $e^{-2a}$.

We choose the standard map \cite{chirikov}, to be our base map $M$ as it  
is a prototypical area-preserving system and is widely
studied in a variety of problems of both theoretical and experimental
interest.  The standard map serves as a test bed for
various models regarding transport \cite{white98}. 
The  map equation is given by,

\begin{eqnarray}
{\bf x}_{n+1} & = & {\bf x}_n + {\bf y}_{n+1} \ ~~~~~~~~~~~ (Mod \  1)\nonumber \\
{\bf y}_{n+1} & = & {\bf y}_n +  \frac{K}{2\pi}\sin( 2\pi{\bf x}_n) \ ~~(Mod \  1).
\label{standard}
\end{eqnarray}

\begin{figure}[! t]
\begin{tabular}{cccc}
\hspace{-0.8cm}
(a)&
 \resizebox{65mm}{65mm}{\includegraphics{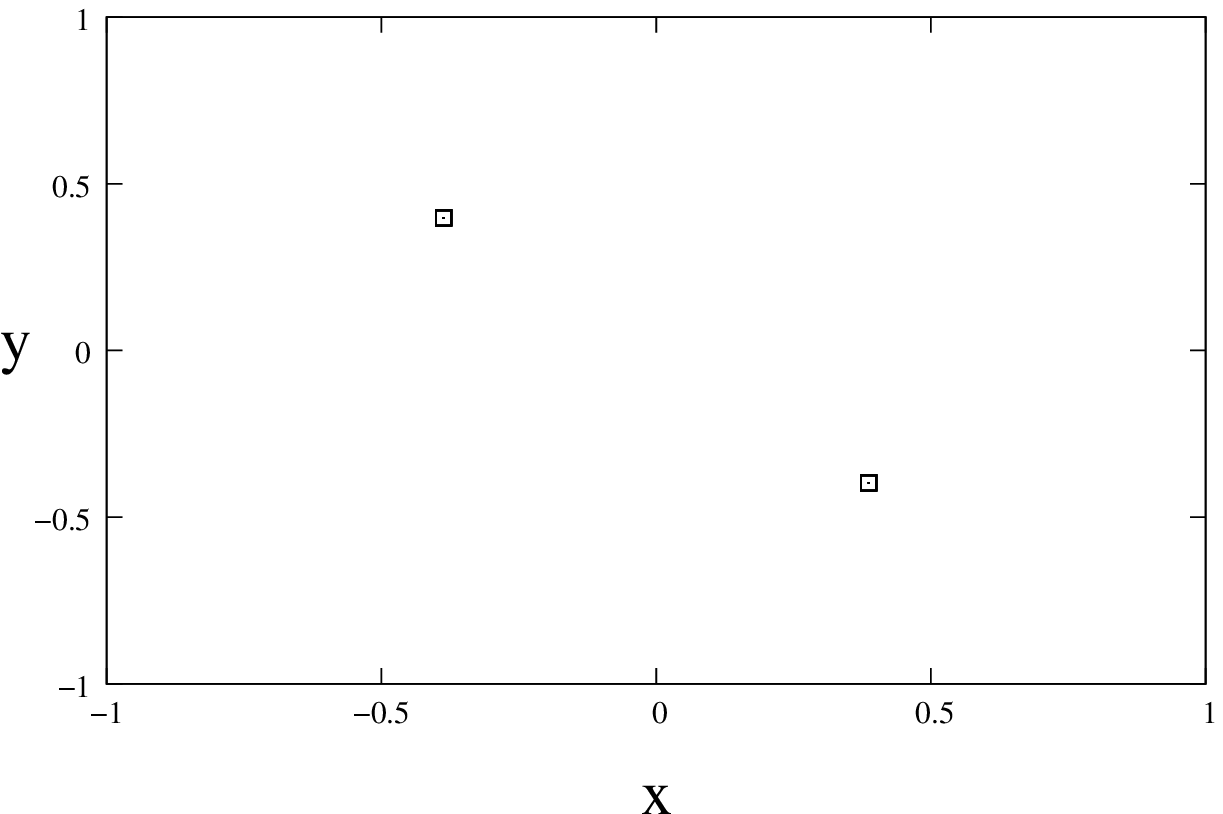}}&
\hspace{0.8cm}
(b)&
\resizebox{65mm}{65mm}{\includegraphics{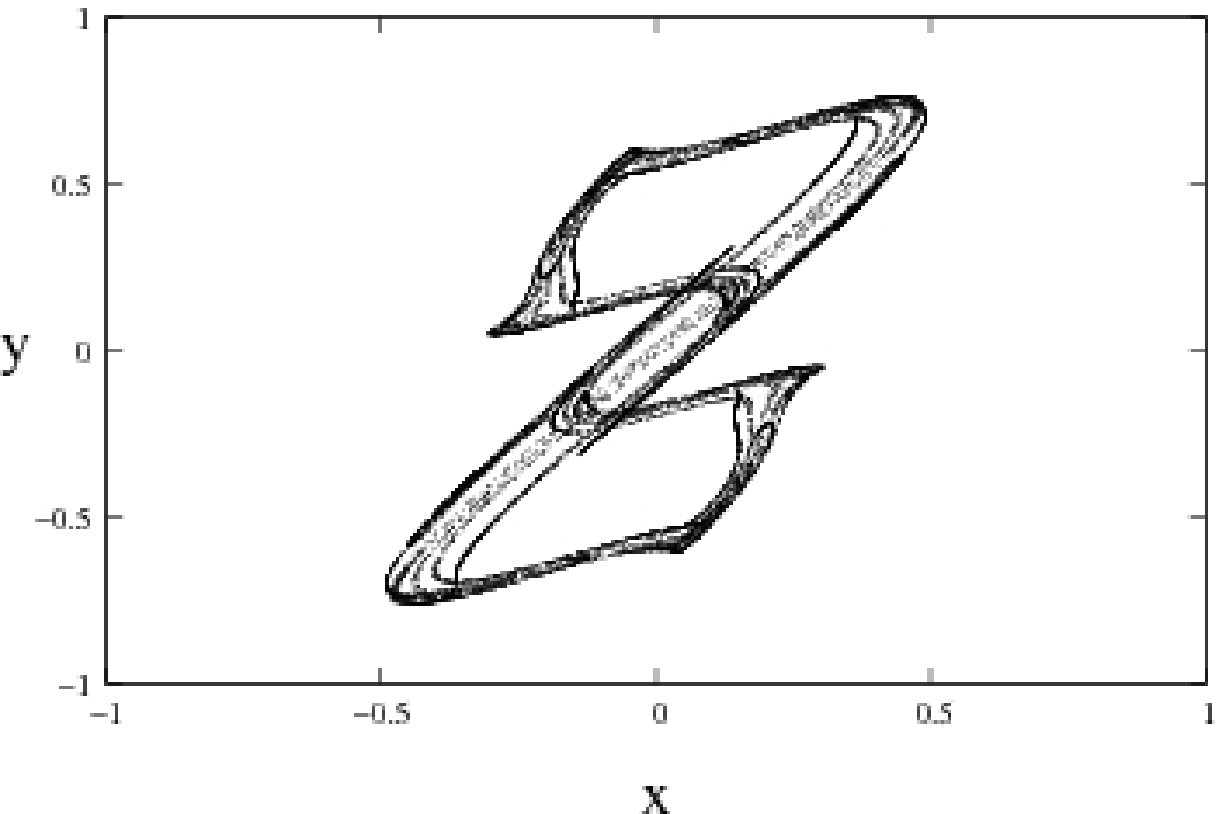}}\\
\end{tabular}                                             

\begin{tabular}{cccc}
\hspace{-0.8cm}
(c)&
 \resizebox{65mm}{65mm}{\includegraphics{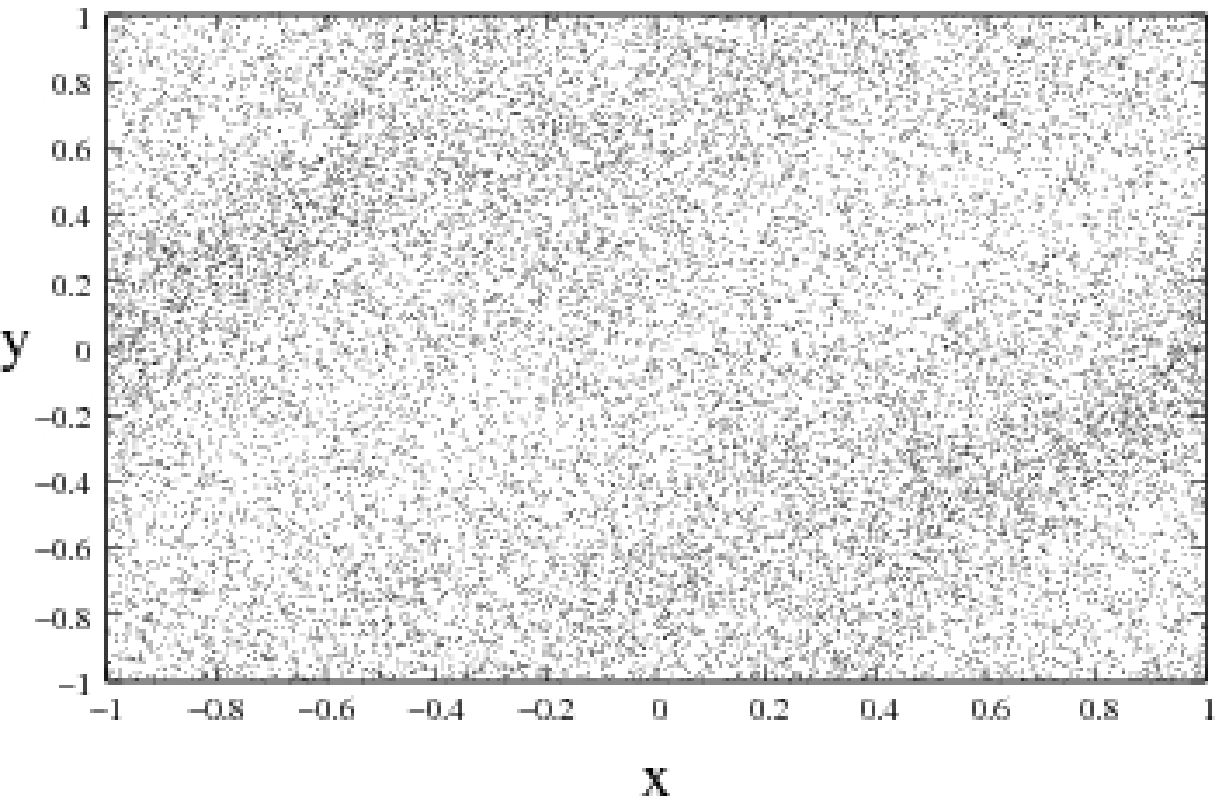}}&
\hspace{0.8cm}
(d)&
\resizebox{65mm}{65mm}{\includegraphics{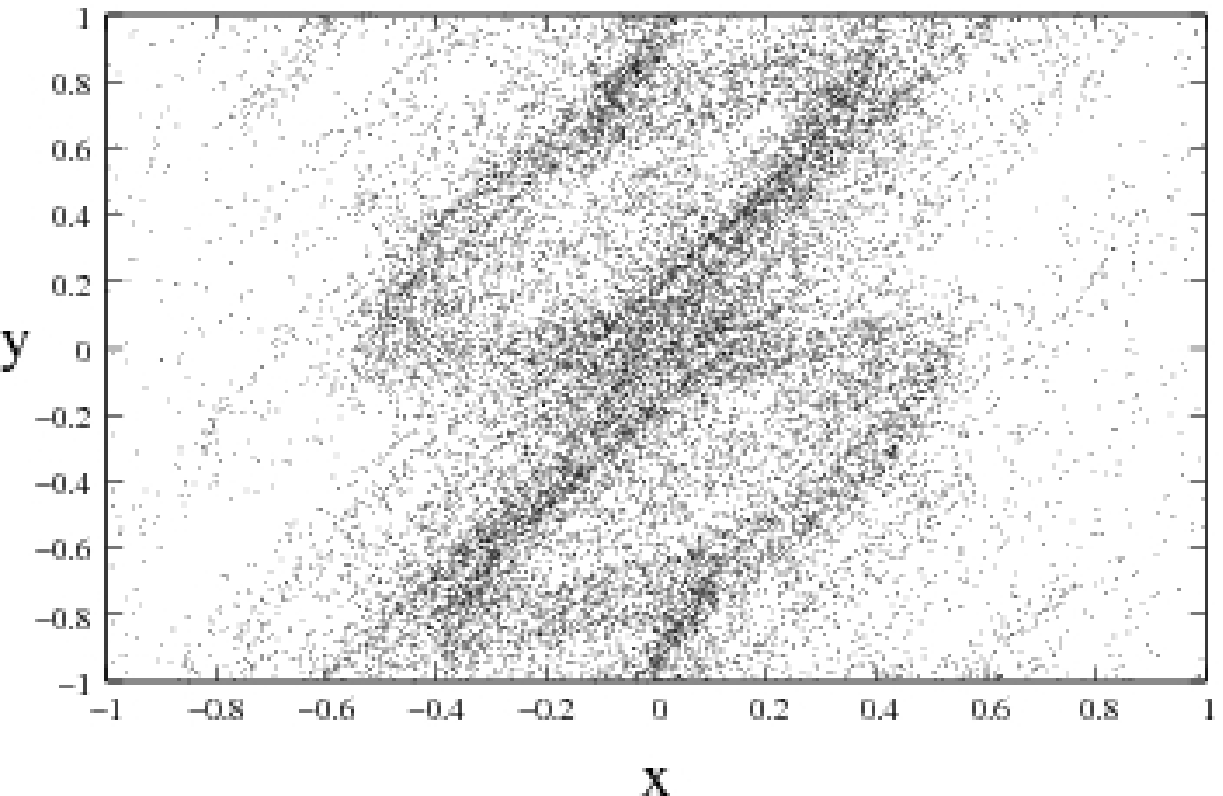}}\\
\end{tabular}                                             
\caption{Phase space plots showing four different 
dynamical regimes of the embedding map.Here $K=2.0$ is fixed.
(a) periodic trajectories ($\alpha=1.5, \gamma=0.7$), (b) chaotic structure
 ($\alpha=0.2$,$\gamma=0.9$), (c) full mixing ($\alpha=2.0$,$\gamma=0.9$)and 
(d) chaotic structure in a mixing background ($\alpha=0.7, \gamma=0.18$), \label{classes}}     
\end{figure}

The parameter $K$ controls the chaoticity of the map. 
This map is taken to 
be the  base map ${\bf M(x}_n)$ in Eqn. (3). The
phase space of the base standard map for $K=2.0$ is plotted in Fig.
\ref{phsp-std} in the region $x \in  [-1,1], y \in   [-1,1]$. 
The particle dynamics is governed by the 4 dimensional map ,
\begin{eqnarray}
x_{n+1} & = & x_{n} + \frac{K}{2\pi}\sin(2\pi y_{n}) + \delta_{n}^{x}  \nonumber\\
y_{n+1} & = & x_{n} + y_{n} + \frac{K}{2\pi}\sin(2\pi y_{n}) + \delta_{n}^{y} \nonumber\\
\delta_{n+1}^{x} & = & e^{-\gamma}[\alpha x_{n+1}-(x_{n+1} - \delta_{n}^{x})] \nonumber\\
\delta_{n+1}^{y} & = & e^{-\gamma}[\alpha y_{n+1}-(y_{n+1} - 	\delta_{n}^{y})]
\label{bailstd}
\end{eqnarray}
It is clear that this  4-dimensional map is  invertible and dissipative.
The  embedding map has 3 parameters $K,\alpha,$ and  $\gamma$.   
We plot the phase space portrait of the system evolved with random
initial conditions at $K=2.0$. The phase space plot 
of the standard map at the same value of $K$ in Fig. \ref{phsp-std} shows
islands and  chaotic regions separated by invariant tori which
constitute barriers to transport.
We plot the embedding map for the aerosol and the bubble cases,
in Fig. \ref{classes}. Unlike earlier studies which claim that only
the aerosols are able to breach the invariant tori, we observe that 
that both the aerosols and bubbles have broken the barrier posed 
by the invariant curve and have targeted the  islands forming periodic
trajectories and structures. Thus, it is clear that 
the dissipation parameter $\gamma$ plays a vital role in
the behaviour of both kinds of particles, the aerosols and the bubbles.
Depending on particular values of $\gamma$ and $\alpha$ both 
these particles can form clusters or undergo mixing in the phase space.
Therefore in order to identify the regimes in which clustering
or mixing can take place, it is necessary to study the full phase
diagram in the parameter space.

\section{The Phase Diagram}

\begin{figure}[! t]
\resizebox{100mm}{80mm}{\includegraphics{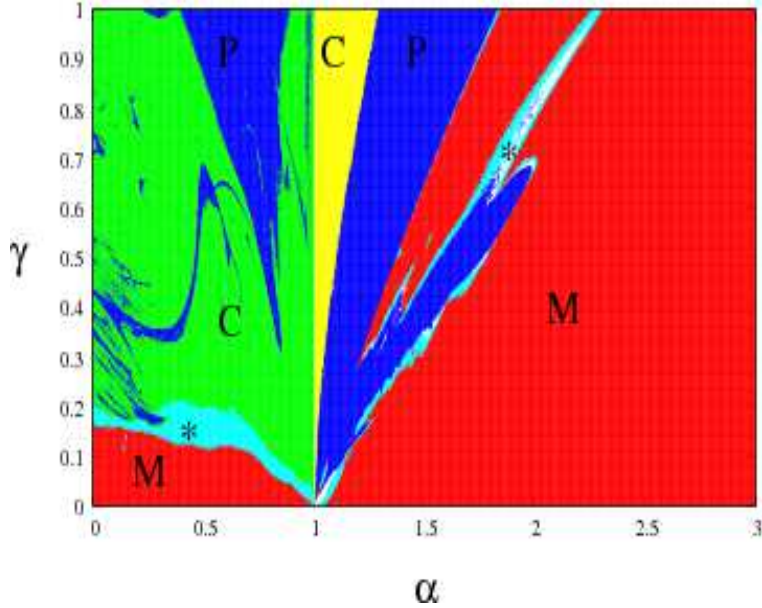}}\\
\caption{(Color Online) The phase diagram.
The regimes with periodic structures are marked
with P (blue). The regimes with chaotic structures are marked with C (green
 for aerosols and yellow for bubbles). The regimes with structures
in a mixing background are marked with a `*'(light blue).
The regimes with full mixing are marked with M (red). \label{lyap-pd}}
\end{figure}

The  phase diagram of the system can be constructed using the following 
criteria. The regimes where the  largest Lyapunov 
characteristic exponent (LCE) $\lambda_{max}>0$ are regions of chaos and  the
regimes where $\lambda_{max}<0$ are regions of periodic behaviour. This property
is used to identify the periodic regimes (marked in blue) in the phase
diagram (see Fig. \ref{lyap-pd}). A typical  phase space plot
in the periodic regime is shown in Fig. \ref{classes} (a). For the calculation
of the Lyapunov exponent, the first 5000 iterates are taken as transients 
and the next 1000 iterates are stored, for 100 random initial 
conditions, uniformly spread in the phase space. We use the 
 Gram-Schmidt reorthonormalization (GSR) procedure for the calculation 
of all the Lyapunov exponents of the embedded map \cite{wolf}.

\begin{figure}[! t]
\resizebox{100mm}{80mm}{\includegraphics{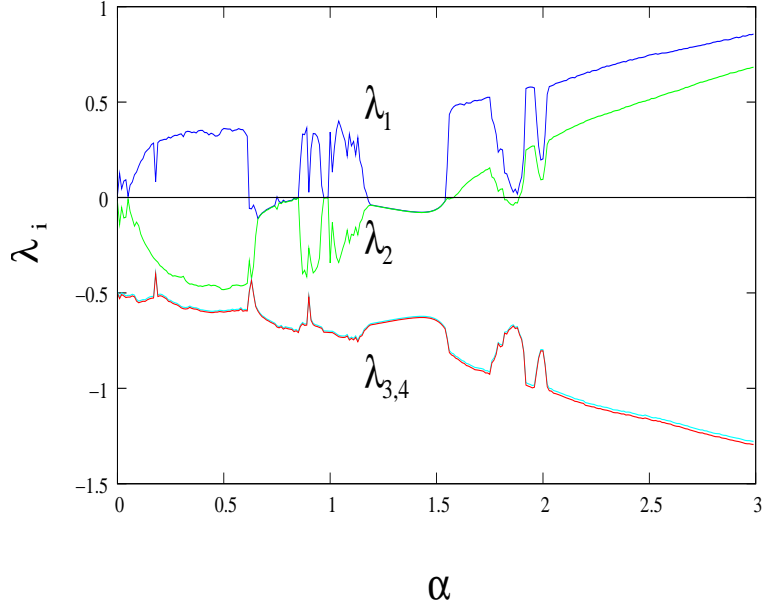}}\\
\caption{(Color Online) Four Lyapunov exponents of the embedding standard
map; $K=2.0$ and $\gamma=0.7$. Intervals for which  $\lambda_{1,2,3,4}<0$ have periodic orbits, and $\lambda_{2,3,4}<0$ have
chaotic structures and $\lambda_{1,2}>0$ have mixing regimes in the phase space.\label{4lp}}
\end{figure}

In the chaotic regimes, further classification can be carried out based on a scheme
which uses the signs of the other Lyapunov exponents of the system  and the 
occupation densities  of the trajectories in the phase space. 
Fig. \ref{4lp} shows all the Lyapunov exponents of the 
system for the dissipation $\gamma=0.7$ and $\alpha=0 \to 3$. 
Chaotic structures in the phase space are found in 
the regimes where $\lambda_{2,3,4}<0$ 
while the largest LCE  $\lambda_{1}$ (i.e.,$\lambda_{max}$) is greater than $0$. 
These are marked in green in the aerosol regions and
yellow in the bubble regions in the phase diagram. For the regimes where both 
$\lambda_{1}$ and $\lambda_{2}$ greater than $0$, 
mixing is seen in the phase space. 

Fig. \ref{classes} shows the phase space plots seen in  three different dynamical regimes. Fig. \ref{classes}(a)
shows the periodic orbits in the bubble regimes
($\alpha=1.5,\gamma=0.7$). Here, both
$\lambda_1$ and $\lambda_2$ are less than $0$. Chaotic structures are
shown in Fig. \ref{classes}(b) in the aerosol regime ($\alpha=0.2,\gamma=0.9$), where
$\lambda_1 > 0$  and $\lambda_2 < 0$. The Figs. \ref{classes} (c),(d) show the 
mixing regimes for the bubble ($\alpha=2,\gamma=0.9$) and the aerosol ($\alpha=0.7,\gamma=0.18$)
regimes respectively. In both the cases for the mixing regime, $\lambda_1$ and $\lambda_2$ are
greater than $0$. We also note that the Lyapunov exponents $\lambda_3$ and $\lambda_4$ are 
less than $0$ in all the three regimes. 

We see that the criterion based on the signs of Lyapunov exponents 
cannot distinguish between the phase space plots showing mixing regime (Fig. \ref{classes} (c))
and the chaotic structure in a mixing background (Fig. \ref{classes} (d)). Therefore we 
use the occupation  densities to distinguish between these two
cases.

The occupation  densities in phase space can be calculated by covering the phase space by a 
$100 \times 100$ mesh, and counting the  number of boxes accessed by the iterates.
The initial conditions and the transients are same as those that
were used for calculating the Lyapunov exponent. 
The regions of parameter space where the iterates access more than
$99 \%$ of the phase space grid turn out to be  fully mixing regions 
(marked in red in Fig. \ref{lyap-pd}), 
with no remnant of any chaotic structure seen
in the phase space plot, and have been marked with an M.
The phase space plot of a mixing regime is seen in Fig. \ref{classes}.
The regions where the iterates  access  more than $90\%$ and less than $99\%$ of the boxes in the phase
space are seen to contain  a  chaotic structure in a mixing background 
(marked with a `*' - light blue  in Fig. \ref{lyap-pd}). Fig. \ref{classes} (d) 
shows the phase space plot of a chaotic structure in  a mixing background.

Periodic structures are seen inside tongues in the
parameter space in the aerosol as well as bubble regimes. 
Thus, the clustering or preferential concentration of inertial 
particles can be seen in both the aerosol and bubble regimes. 
Chaotic structures can be seen outside the tongues 
in the aerosol regime and in the border  regions of the
tongues in the bubble regime. The aerosol part ($0 < \alpha < 1$) of the phase
diagram  has a small region of mixing
 while the bubble part ($1 < \alpha < 3$) is dominated by the mixing regime. 
 In recent work we reported \cite{nirmal} that
crisis induced intermittency is seen  at some parameters in the aerosol
regime in the phase diagram.  
Here the characteristic time between bursts scales algebraically with a function
of  the dissipation parameter $\gamma$, with the power law 
exponent to be $\beta=-1/3$. 

%------------------------------------------------------------
%-----------------------------------------------------------------------------------------------

\section{Conclusion}

The present work discusses the Lagrangian dynamics
of passive scalars of finite size in an incompressible
two-dimensional flow for situations where 
the particle density can differ from that of the fluid. 
The motion of the advected particles is represented by a
dissipative embedding map 
with the  area preserving standard map as the base map. 

The embedded  standard  map  has three parameters,
the nonlinearity $K$, the dissipation parameter $\gamma$ 
and the mass ratio parameter $\alpha$ which measures the extent to 
which the density of the particles differs from that of the fluid. 
The phase diagram of the system in the $\alpha-\gamma$ parameter space 
shows very rich structure with three types of dynamical behaviours -  
periodic orbits, chaotic structures, and mixed regimes which can be
partially or fully mixed. The Lyapunov
characteristic exponents (LCE) of the $4-d$ map are  used 
to demarcate these dynamical regimes in the 
phase diagram. We observed that both aerols and bubbles can breach the
invariant curves of the base map and form clusters, depending on the
dissipation parameter of the system.

Our study can have implications for the preferential concentration 
of inertial particles in flows, as well as for the targeting 
of periodic structures. Earlier results for such systems indicated that bubbles could
breach elliptic islands and target structures, whereas aerosols could
not. Our results indicate that 
both aerosols and bubbles can breach the 
invariant curves of the base map and form clusters, depending on the 
dissipation parameter of the system. Mixing regimes can also exist for
both aerosols and bubbles.
Thus the dissipation parameter $\gamma$ plays
as crucial role in clustering and mixing  as does the density differential parameter $\alpha$,
and the examination of the full phase diagram is necessary to draw
conclusions about the parameter regimes where targeting and breaching
can occur.  
These results can have implications for the dynamics of impurities in
diverse application contexts, e.g. that of the dispersion of pollutants 
in the atmosphere, flows with suspended micro-structures, coagulation of material
particles in flows, catalytic chemical reactions and 
the plankton population in oceans.

\end{document}